\documentclass[aps,prl,reprint,amsmath,amssymb,showpacs,superscriptaddress]{revtex4-1}

\usepackage{color}

\usepackage{graphicx}
\usepackage{dcolumn}
\usepackage{bm}
\usepackage{braket}%
\usepackage{natbib}
\usepackage{hyperref}
\hypersetup{%
 colorlinks=true, citecolor=blue, filecolor=blue, linkcolor=blue, urlcolor=blue}
\usepackage{subfigure} 

\begin{document}

\preprint{Europhysics Letters}

\title{Central-moments-based lattice Boltzmann scheme \\
for coupled Cahn-Hilliard--Navier-Stokes equations}



\author{Alessandro De Rosis}
\email{derosis.alessandro@icloud.com}

\author{Shimpei Saito}

\author{Akiko Kaneko}

\author{Yutaka Abe}

\date{\today}

\begin{abstract}
	In this paper, we propose a lattice Boltzmann (LB) model to solve the coupled Cahn-Hilliard-Navier-Stokes equations. Differently from previous efforts, the LB equation for the fluid velocity is decomposed in a space of non-orthogonal central moments where the surface tension force is inserted directly in the equilibrium state. The present scheme is validated against well-consolidated benchmark tests, showing very good accuracy. Moreover, it outperforms the BGK approach in terms of stability, as our algorithm allows us to simulate a very large range of viscosity contrast.
\end{abstract}


\maketitle

\begin{figure}[htpb]
\centering
\subfigure{\includegraphics[width=0.30\textwidth]{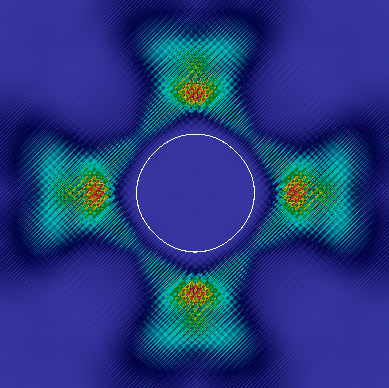}\label{Figure1}}
\subfigure{\includegraphics[width=0.30\textwidth]{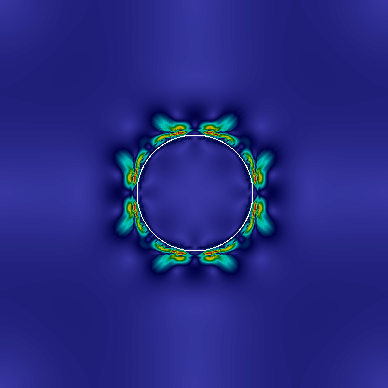}\label{Figure2}}
\caption{Map of the velocity field by (top) a double-BGK run at $\zeta=3$ and (bottom) a hybrid CMS-BGK run at $\zeta=100$. The white line denotes the interface between the two phases. Very noisy velocities are evident in the top panel, corresponding to a rapid onset of instability. Conversely, the map appears significantly smoother in the bottom panel.}
\end{figure}
\begin{figure}[htpb]
\centering
\includegraphics[scale=1]{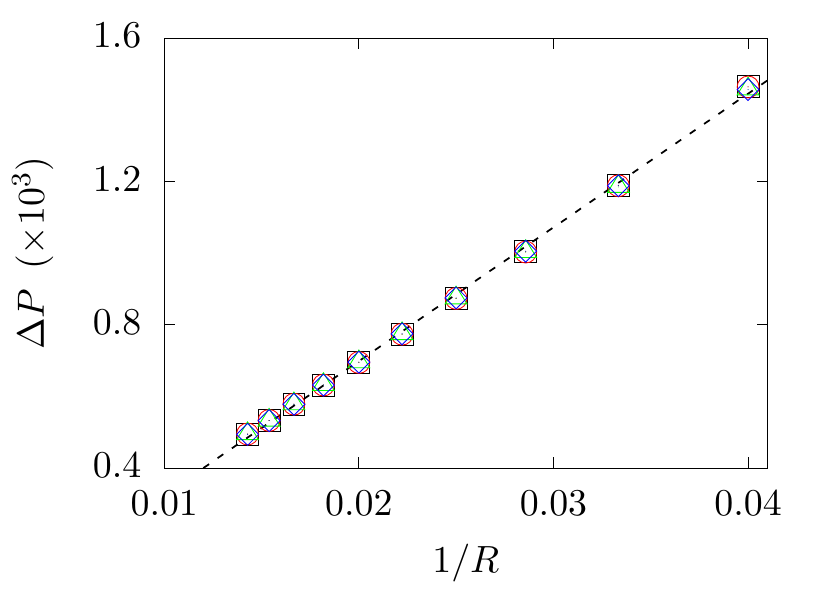}
\caption{Laplace's law: pressure jump vs bubble radius for different values of the viscosity ratio, \textit{i.e.} $\zeta=2$ (black squares), 10 (red circles), 100 (green triangles) and 1000 (blue diamonds). Data are fitted by a dashed black line whose slope is 0.03717, showing a very good agreement with the analytical predicition of 0.03771.}
\label{Figure3}
\end{figure}
\begin{figure}[htpb]
\centering
\includegraphics[scale=1]{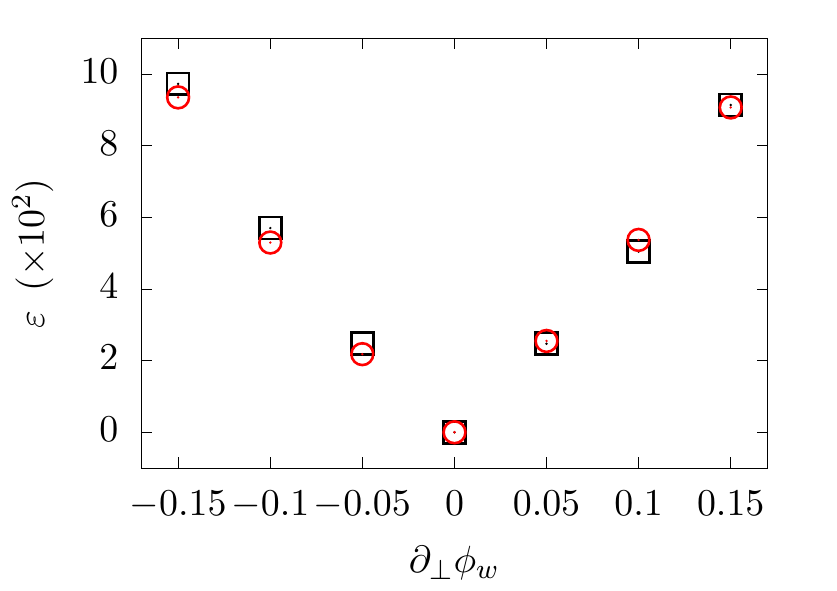}
\caption{Contact angles: relative error between analytical predictions and numerical results by BGK (black squares) and present model (red circles).}
\label{Figure4}
\end{figure}
\begin{figure}[htbp!]
\centering
\includegraphics[width=0.3\textwidth]{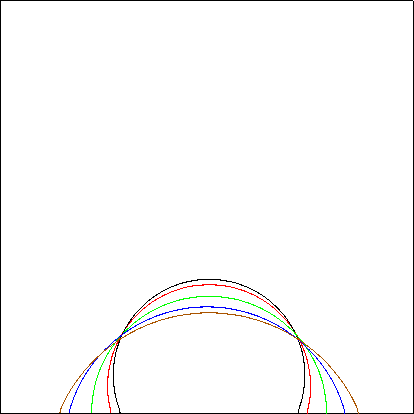}
\caption{Contact angles: contour plot of the interface shape for different values of the wall gradient, \textit{i.e.} -0.15 (black), -0.1 (red), 0 (green), 0.1 (blue) and 0.15 (brown).}
\label{Figure5}
\end{figure}
\begin{figure*}[htbp!]
\centering
\subfigure{\includegraphics[width=0.18\textwidth]{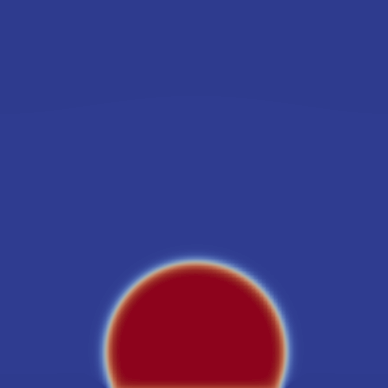}}
\subfigure{\includegraphics[width=0.18\textwidth]{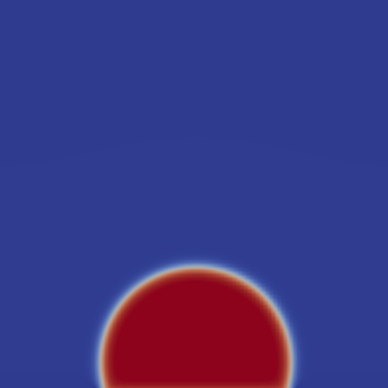}}
\subfigure{\includegraphics[width=0.18\textwidth]{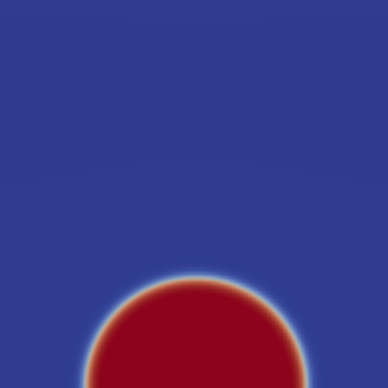}}
\subfigure{\includegraphics[width=0.18\textwidth]{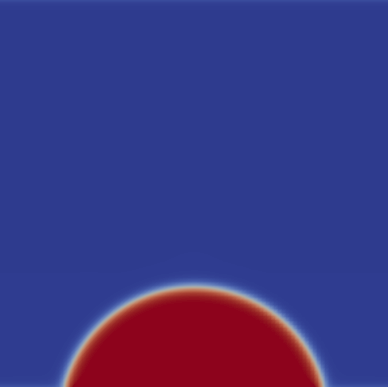}}
\subfigure{\includegraphics[width=0.18\textwidth]{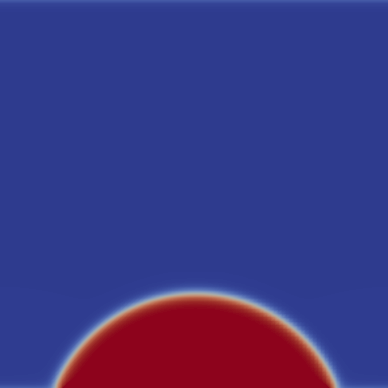}}
\caption{Contact angles: map of the phase field for different values of the wall gradient, \textit{i.e.} -0.15, -0.1, 0, 0. and 0.15 (from left to right).}
\label{Figure6}
\end{figure*}
\begin{figure}[htpb]
\centering
\includegraphics[scale=1]{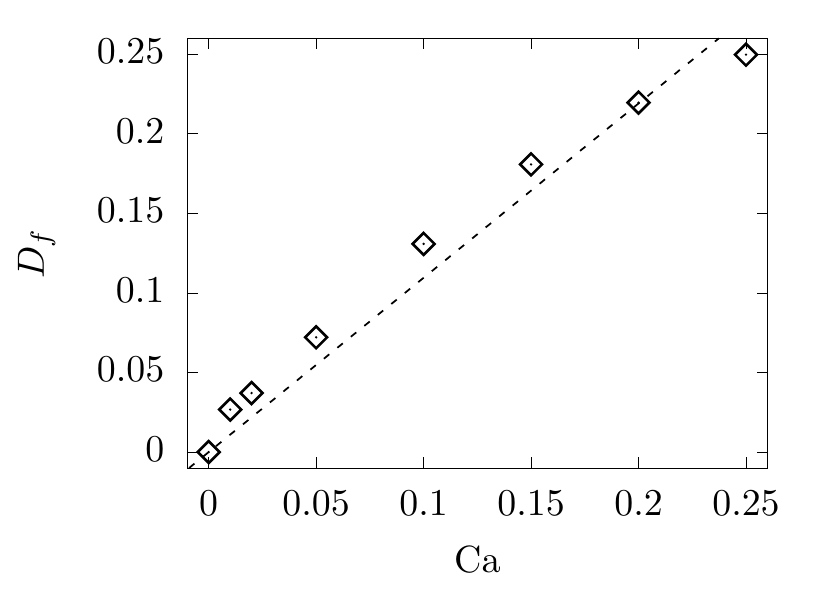}
\caption{Droplet in a shear flow: deformation of the droplet as a function of the capillarity number. Data are fitted by a dashed black line whose slope is 1.09536.}
\label{Figure7}
\end{figure}
\begin{figure}[htbp!]
\centering
\includegraphics[width=0.3\textwidth]{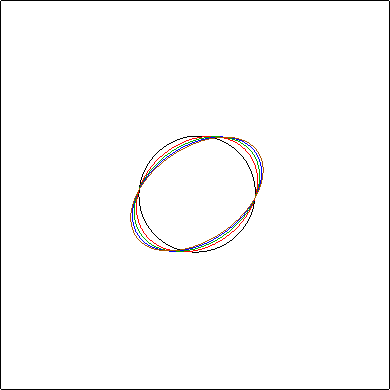}
\caption{Droplet in a shear flow: contour plot of the interface shape for different values of the capillarity number, \textit{i.e.} 0.01 (black), 0.1 (red), 0.15 (green), 0.2 (blue) and 0.25 (brown).}
\label{Figure8}
\end{figure}
\begin{figure}[htbp!]
\centering
\subfigure{\includegraphics[width=0.45\textwidth]{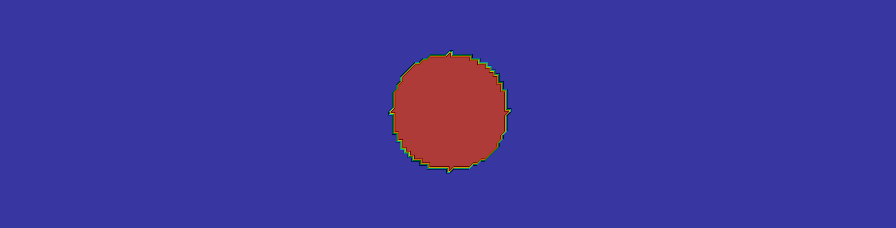}}\\
\subfigure{\includegraphics[width=0.45\textwidth]{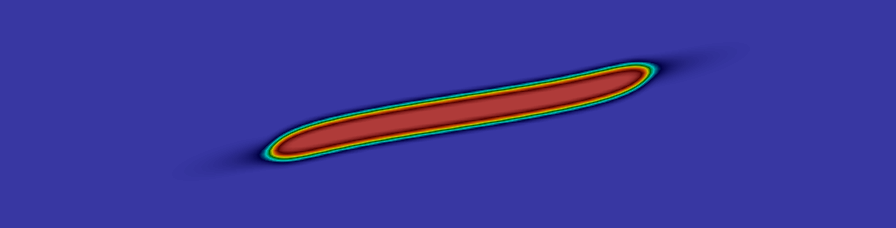}}
\subfigure{\includegraphics[width=0.45\textwidth]{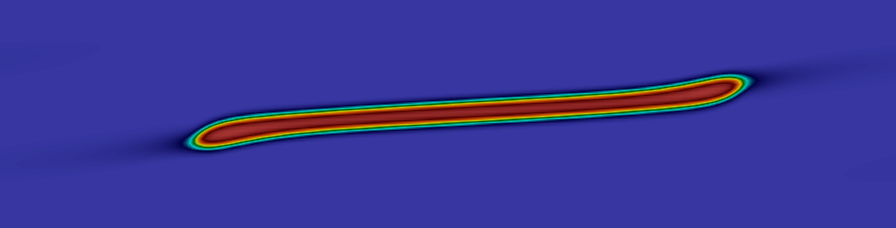}}
\subfigure{\includegraphics[width=0.45\textwidth]{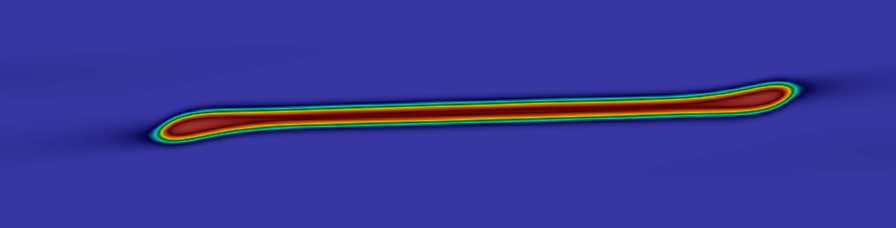}}
\subfigure{\includegraphics[width=0.45\textwidth]{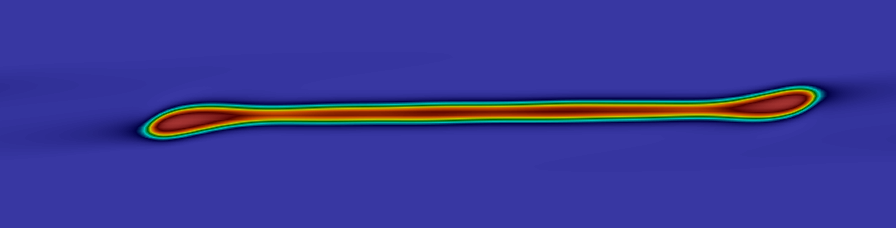}}
\subfigure{\includegraphics[width=0.45\textwidth]{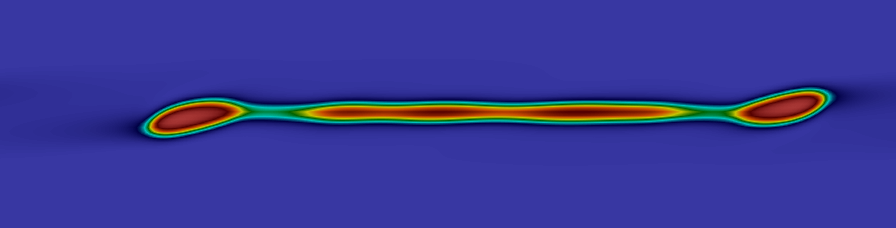}}
\subfigure{\includegraphics[width=0.45\textwidth]{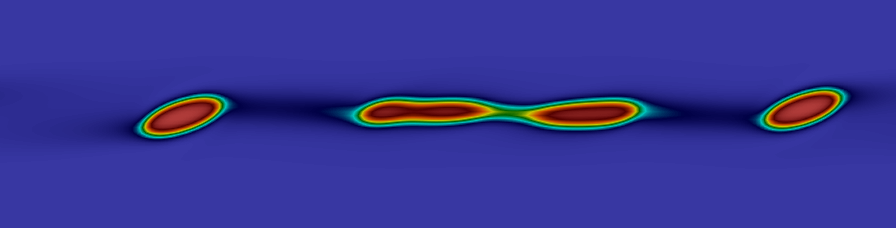}}
\caption{Droplet in a shear flow: map of the phase field at salient time instants, \textit{i.e.} t=0 (first), $5t^{\dagger}$ (second), $10t^{\dagger}$ (third), $15t^{\dagger}$ (fourth), $20t^{\dagger}$ (fifth), $25t^{\dagger}$ (sixth) and $30t^{\dagger}$ (seventh).}
\label{Figure10}
\end{figure}

Phase changes are ubiquitous in nature. Melting ice, freezing water into ice cubes and boiling are popular examples routinely experienced, as well as the formation of snow into the clouds. Interestingly, phase transitions play an important role in several industrial areas. Common applications involve cooling devices, ranging from medicine/vaccine storage to fresh product transportation. Moreover, the design of battery thermal management systems for electric vehicles is a challenging problem involving phase change materials. A deep understanding of the laws governing phase transitions may significantly improve the design process of such industrial systems.\\
\indent Originated by van der Waals~\cite{van1979thermodynamic} and Cahn-Hilliard~\cite{cahn1958free, cahn1959free}, the so-called phase-field theory introduces a variable $\phi$ (also known as order parameter) describing the transition of matter between different states or phases. The thermodynamic behavior of a binary two-phase fluid can be written as a function of a Landau free energy functional $\displaystyle E = \int \left( \psi + \frac{\gamma}{2} | \boldsymbol \nabla \phi  |^2   \right) \mathrm{d}V + \int  \xi \phi \mathrm{d}S$~\cite{penrose1990thermodynamically}, where $V$ and $S$ are the volume and the surface of the system, respectively, $\gamma$ is related to the surface tension between the two phases through $\sigma = \sqrt{8 \gamma a/9}$, $a$ being a constant, and $\xi$ controls the interface thickness. The bulk free energy density is $\displaystyle \psi = \frac{1}{3} \rho \mathrm{ln}\rho + a \left( -\frac{1}{2}\phi^2 + \frac{1}{4} \phi^4   \right)$, that corresponds to a binary separation into two phases with $\phi = \pm 1$. By performing the variation of the free energy with respect to the order parameter, it is possible to obtain the chemical potential $\displaystyle \mu = \frac{\partial E(\phi)}{\partial \phi} = a\left(-\phi + \phi^3 \right) - \gamma \boldsymbol \Delta \phi$~\cite{badalassi2003computation}. The problem is governed by the 
following set of equations:
\begin{eqnarray}\label{CH}
\partial_t \phi + \bm{v} \cdot \boldsymbol \nabla \phi &=& M \boldsymbol \Delta \mu, \\
\boldsymbol \nabla \cdot \bm{v} &=& 0,\\
\rho \left(  \partial_t \bm{v} +\bm{v}  \cdot \boldsymbol \nabla \bm{v} \right) &=& -\boldsymbol \nabla \bm{P} + \nu \boldsymbol \Delta \bm{v},\label{NS}
\end{eqnarray}
where $t$ is the time, $\bm{v}$ is the fluid velocity, $M$ is a mobility coefficient, $\rho$ is the density and $\nu$ is the kinematic viscosity. Eq.~(\ref{CH}) is the convective Cahn-Hilliard equation and the other two represent the incompressible Navier-Stokes ones, governing the behavior of a density-matched fluid when $\rho$ is constant~\cite{jacqmin1999calculation,kim2005continuous}. The pressure tensor is defined as $\displaystyle P_{\alpha \beta} = \left(  p_0 + \gamma \phi \boldsymbol \Delta \phi - \frac{\gamma}{2} | \boldsymbol \nabla \phi  |^2 \right) \delta_{\alpha \beta} + \gamma \partial_{\alpha} \phi \partial_{\beta} \phi$, where $\alpha$ and $\beta$ span the Eulerian basis, $\delta_{\alpha \beta}$ is the Kronecker operator and $p_0$ is the bulk pressure.\\
\indent Eqs.~(\ref{CH}-\ref{NS}) can be solved within the framework of the lattice Boltzmann method~\cite{benzi1992lattice, he1997theory, chen1998lattice}. He \textit{et al.}~\cite{he1999lattice} proposed the first phase-field LB model for incompressible multiphase flows, where an index function was introduced to capture the evolution of the interface between two phases. Ref.~\cite{briant2004lattice} introduced a modified equilibrium distribution function for the order parameter. Significant efforts have been put into the performance improvement of phase-field LB model, especially for high-density-ratio multiphase flows by using the projection method~\cite{Inamuro2004}, the stable discretization~\cite{Lee2005}, the modified LB equation~\cite{Zheng2006}, and the entropic method~\cite{MazloomiM2015}. Several other models proved to be able to recover correctly the Cahn-Hilliard equation~\cite{zheng2005lattice, lee2010lattice, zu2013phase, zheng2015lattice}. The interested reader can refer to~\cite{li2016lattice} (and references therein) for a comprehensive review of the phase-field LB modeling.\\
\indent Let us consider a two-dimensional Eulerian basis $\bm{x} = [x,y]$. In order to predict the behavior of a binary fluid, two groups of particle distribution functions (or populations) are considered. The former, $| f_{i}\rangle = \left[ f_0,\, f_1,\, f_2,\, f_3,\, f_4,\, f_5,\, f_6,\, f_7,\, f_8    \right]^{\mathrm{{T}}}$, controls the velocity field, while the latter, $| g_{i}\rangle = \left[ g_0,\, g_1,\, g_2,\, g_3,\, g_4,\, g_5,\, g_6,\, g_7,\, g_8    \right]^{\mathrm{{T}}}$, monitors the evolution of the order parameter. Notice that $| \bullet \rangle$ denotes a column vector and the superscript $\top$ indicates the transpose operator. In the D2Q9 model~\cite{SucciBook}, populations move on a fixed Cartesian square lattice along the generic link $i=0 \ldots 8$ with velocity $\mathbf{c}_i=[| c_{xi}\rangle ,\, | c_{yi}\rangle]$ defined as $\displaystyle | c_{xi}\rangle = \left[0,\,1,\,0,\,-1,\,0,\,1,\,-1,\,-1,\,1      \right]^{\top}$ and $\displaystyle | c_{yi}\rangle = \left[0,\,0,\,1,\,0,\,-1,\,1,\,1,\,-1,\,-1      \right]^{\top}$. The LB equations reads as follows:
\begin{eqnarray}
f_i(\bm{x}+\Delta t \bm{c}_i,t+\Delta t) &=& f_i^{\star}(\bm{x},t),  \label{BGK_Velo}\\
g_i(\bm{x}+\Delta t \bm{c}_i,t+\Delta t) &=& g_i^{\star}(\bm{x},t). \label{BGK_Phase}
\end{eqnarray}
where the time step is $\Delta t=1$ and the superscript $\star$ represents the so-called post-collision state. Within the BGK approximation~\cite{bhatnagar1954model}, populations relax to equilibrium states, \textit{i.e.} 
\begin{eqnarray}
f_i^{\star}(\bm{x},t) &=& f_i(\bm{x},t) + \omega_{\rho} \left[ f_i^{eq}(\bm{x},t) - f_i(\bm{x},t) \right]\label{BGK_eq_Velo}\\
g_i^{\star}(\bm{x},t) &=& g_i(\bm{x},t) + \omega_{\phi} \left[ g_i^{eq}(\bm{x},t) - g_i(\bm{x},t) \right],\label{BGK_eq_Phase}
\end{eqnarray}
where ~\cite{pooley2008eliminating}:
\begin{equation}
\begin{split}
f_i^{eq} &= w_i \left[ c_s^{-2} \tilde{p} + \rho \left( \frac{\mathbf{c}_i \cdot  \bm{v}}{c_s^2} +  \frac{\left( \mathbf{c}_i \cdot  \bm{v} \right)^2}{2c_s^4} - \frac{\bm{v}^2}{2c_s^2} \right) \right]\\
 &+ \gamma \left[ w_i^{xx} (\partial_x \phi)^2+ w_i^{yy} (\partial_y \phi)^2  + w_i^{xy} \partial_x \phi \partial_y \phi \right],
\end{split}
\end{equation}
\begin{equation}
g_i^{eq} = w_i \left[\Gamma \mu + \phi \left( \frac{\mathbf{c}_i \cdot  \bm{v}}{c_s^2} +  \frac{\left( \mathbf{c}_i \cdot  \bm{v} \right)^2}{2c_s^4} - \frac{\bm{v}^2}{2c_s^2} \right)   \right],
\end{equation}
for $i=1\ldots 8$ and
\begin{equation}
f_0^{eq} = \rho - \sum_{i=1}^8 f_i^{eq}, \qquad g_0^{eq} = \phi - \sum_{i=1}^8 g_i^{eq}.
\end{equation}
The effect of the surface tension is inserted directly in the equilibrium through the weighting factors
\begin{eqnarray}
w_i^{xx} &=& \left[0, \frac{1}{3}, \frac{-1}{6}, \frac{1}{3}, \frac{-1}{6}, \frac{-1}{24}, \frac{-1}{24}, \frac{-1}{24}, \frac{-1}{24} \right], \nonumber \\
w_i^{yy} &=& \left[0, \frac{-1}{6}, \frac{1}{3}, \frac{-1}{6}, \frac{-1}{3}, \frac{-1}{24}, \frac{-1}{24}, \frac{-1}{24}, \frac{-1}{24} \right], \nonumber \\
w_i^{xy} &=& \left[0,\, 0,\, 0,\, 0,\, 0,\, \frac{1}{4},\, \frac{-1}{4},\, \frac{1}{4},\, \frac{-1}{4} \right].
\end{eqnarray}
Gradient and laplacian operators are computed by applying the following stencils:
\begin{eqnarray}
\partial_x &=& \left[0,\, \frac{4}{12},\, 0,\, \frac{-4}{12},\, 0,\, \frac{1}{12},\, \frac{-1}{12},\, \frac{-1}{12},\, \frac{1}{12} \right],\nonumber \\
\partial_y &=& \left[0,\, 0,\, \frac{4}{12},\, 0,\, \frac{-4}{12},\, \frac{1}{12},\, \frac{1}{12},\, \frac{-1}{12},\, \frac{-1}{12} \right],\nonumber \\
\boldsymbol \Delta &=& \left[\frac{-20}{6},\, \frac{4}{6},\, \frac{4}{6},\, \frac{4}{6},\, \frac{4}{6},\, \frac{1}{6},\, \frac{1}{6},\, \frac{1}{6},\, \frac{1}{6} \right].
\end{eqnarray}
The bulk pressure is
\begin{equation}
p_0 = \rho c_s^2 + a \left(  -\frac{1}{2} \phi^2 + \frac{3}{4} \phi^4 \right)
\end{equation}
and
\begin{equation}
\tilde{p} = p_0 - \gamma \phi \boldsymbol \Delta \phi,
\end{equation}
$c_s = 1/ \sqrt{3}$ being the lattice sound speed. The weights $w_i$ are $w_0=4/9$, $w_{1 \ldots 4} = 1/9$, $w_{5 \ldots 8}=1/36$. The solution procedure of Eqs.~({\ref{BGK_Velo}}-{\ref{BGK_Phase}}) based on the collision process in Eqs.~(\ref{BGK_eq_Velo}-\ref{BGK_eq_Phase}) can be viewed as a double-population double-BGK approach. The relaxation frequency $\omega_{\phi}$ is related to mobility as $\displaystyle M=  \left( \frac{1}{\omega_{\phi}}  -\frac{1}{2}\right) \Gamma c_s^2$, $\Gamma$ being a free parameter. The frequency $\omega_{\rho}$ is computed as $\displaystyle \omega_{\rho} = \omega_{g} + \frac{1}{2}\left(\phi+1 \right)\left(\omega_{l}-\omega_{g} \right)$, where $\displaystyle \omega_{l}= \left(\frac{\nu_l}{c_s^2}+\frac{1}{2} \right)^{-1}$ and $\displaystyle \omega_{g}= \left(\frac{\nu_g}{c_s^2}+\frac{1}{2} \right)^{-1}$. $\nu_l$ and $\nu_g$ denote the kinematic viscosity of the liquid and gaseous phases, respectively. Macroscopic variables are available through $\displaystyle \rho = \sum_i f_i$, $\displaystyle \bm{v} = \frac{\sum_i f_i \mathbf{c}_i}{\sum_i f_i }$ and $\displaystyle \phi = \sum_i g_i$.\\
\indent Despite the intrisic simplicity of the double-BGK approach, it suffers from numerical instabilities where the ratio between the viscosity of the two phases, namely $\zeta=\nu_l/\nu_g$, increases. Let us consider a domain consisting of $200 \times 200$ lattice points. The phase is set to -1 everywhere, except inside a bubble of radius $R$, where $\phi=1$. The center of the bubble is located at the center of the domain. Let us assume a liquid phase with viscosity $\nu_l=0.1$ and a gaseous one with viscosity $\nu_g =\nu_l/\zeta $. Here and henceforth, the mobility is $M=0.1$ and $\Gamma=1$. Moreover, the fluid is initially at rest and the density is set to 1 everywhere. By setting $\zeta=3$, a double-BGK run is affected by the rise of very high spurious currents, leading to a rapid onset of instability (see Fig.~\ref{Figure1}).
In order to alleviate these deleterious effects, several attempts have been performed within the BGK approximation~\cite{inamuro2000galilean, kalarakis2002galilean, kalarakis2002galilean, pooley2008eliminating} and by the multiple-relaxation-time model~\cite{pooley2008contact, liang2014phase}.\\
\indent In 2006, Geier \textit{et al.}~\cite{geier2006cascaded} proposed a new collision operator baased on the relaxation of central moments which are obtained by shifting the lattice directions by the local fluid velocity. Geier's collision operator is also known as ``cascaded'' due to its particular hierarchical structure, where the post-collision state of a certain moment at a given order depends only on lower order ones. It has been demonstrated that it outperforms the BGK model in terms of stability~\cite{premnath2009incorporating, FLD:FLD4208, geier2015cumulant}. The cascaded scheme has been also used for phase-field modeling \cite{geier2015conservative,fakhari2016simple}. More recently, the adoption of central moments (CMs) has been reinterpreted by introducing a non-orthogonal basis, with the resultant scheme losing the pyramidal topological pattern of the collision process~\cite{derosis2016epl_d2q9, de2017nonorthogonal}. Interestingly, this model entails a very intelligible analytical formulation, it shows an easy practical implementation and it can be easily extended to any lattice velocity space. Among its most compelling features, the procedure outlined in~\cite{derosis2016epl_d2q9, de2017nonorthogonal} can be applied to whatever BGK formulation. For instance, a CMs-based formulation has been proposed for the preconditioned Navier-Stokes equations~\cite{de2017preconditioned} and to solve the shallow waters equations~\cite{de2017centralshallow}.\\
\indent Here, we derive, test and validate a CMs-based model able to recover the solution of the coupled Cahn-Hilliard-Navier-Stokes equations. In particular, we develop a double-population hybrid CMs-BGK approach, where the dynamics of the phase field is predicted by Eqs.~(\ref{BGK_Phase},\ref{BGK_eq_Phase}), while central moments are adopted to evaluate the post-collision state $f_i^{\star}$. The procedure begins by shifting the lattice directions by the local fluid velocity as $\displaystyle | \bar{c}_{xi}\rangle = |c_{xi}-u_x \rangle$ and $\displaystyle | \bar{c}_{yi}\rangle = |c_{yi}-u_y \rangle$~\cite{geier2006cascaded}. Let us use a basis $\displaystyle \bar{\mathcal{T}} = \left[  \bar{T}_0,\, \ldots, \, \bar{T}_i,\, \ldots, \, \bar{T}_8  \right]$, whose components are 
\begin{eqnarray}
|\bar{T}_0\rangle &=& \left[1,1,1,1,1,1,1,1,1   \right]^{\top},\nonumber \\
|\bar{T}_1\rangle &=& | \bar{c}_{xi}\rangle , \qquad  \quad \, \,\, |\bar{T}_2\rangle = | \bar{c}_{yi}\rangle ,\nonumber \\
|\bar{T}_3\rangle  &=& | \bar{c}_{xi}^2+ \bar{c}_{yi}^2\rangle , \quad |\bar{T}_4\rangle = | \bar{c}_{xi}^2- \bar{c}_{yi}^2\rangle ,\nonumber \\
|\bar{T}_5\rangle  &=& | \bar{c}_{xi} \bar{c}_{yi}\rangle, \quad  \quad \, \, |\bar{T}_6\rangle = | \bar{c}_{xi}^2 \bar{c}_{yi} \rangle,\nonumber \\
|\bar{T}_7\rangle  &=& | \bar{c}_{xi}\bar{c}_{yi}^2 \rangle, \quad  \quad \, \, |\bar{T}_8\rangle = | \bar{c}_{xi}^2\bar{c}_{yi}^2 \rangle.
\end{eqnarray}
The matrix $\bar{\mathcal{T}}$ can be interpreted as a transformation one, allowing us to switch from the populations space to the CMs one, and vice versa. Pre-collision CMs are collected as $\displaystyle | k_i \rangle = \left[ k_0,\, \ldots, \, k_i,\, \ldots, \, k_{8}   \right]^{\top}$ and are defined as $\displaystyle | k_i \rangle = \bar{\mathcal{T}}^{\top}  | f_i \rangle$.
\indent Each moment relaxes to an equilibrium state, $k_i^{eq}$, defined through $\displaystyle | k_i^{eq} \rangle = \bar{\mathcal{T}}^{\top}  | f_i^{eq} \rangle$. The resultant expressions of the equilibrium CMs are the following:
\begin{eqnarray}
k_0^{eq} &=& \rho,\nonumber \\
k_1^{eq} &=&  0,\nonumber \\
k_2^{eq} &=&  0,\nonumber \\
k_3^{eq} &=&  2 \tilde{p},\nonumber \\
k_4^{eq} &=&  \gamma \left[ (\partial_x \phi)^2- (\partial_y \phi)^2  \right],\nonumber \\
k_5^{eq} &=& \gamma \partial_x \phi \partial_y \phi,\nonumber \\
k_6^{eq} &=& -\frac{1}{2} v_y k_4^{eq} -2 v_x k_5^{eq} + \rho v_y \left(c_s^2-u_x^2   \right) - \tilde{p}v_y, \nonumber \\
k_7^{eq} &=& \frac{1}{2} v_x k_4^{eq} - 2 v_y k_5^{eq} + \rho v_x \left(c_s^2-u_y^2   \right) - \tilde{p}v_x, \nonumber 
\end{eqnarray}
\begin{equation}
\begin{split}
k_8^{eq} &= \tilde{p}(c_s^2+v_x^2+v_y^2) - \rho c_s^2(v_x^2+v_y^2) + \frac{\rho}{c_s^2} v_x^2 v_y^2 \\
&- \frac{c_s^2}{2}\gamma \boldsymbol \Delta \phi +\frac{1}{2}k_4^{eq} (v_y^2-v_x^2) + 4 v_x v_y k_5^{eq}, \label{discrete}
\end{split}
\end{equation}
where $v_x$ and $v_y$ are the two components of the velocity vector. Notice that if $\phi=0$ the equilibrium state to collapse into the one of the sole Navier-Stokes equations~\cite{derosis2016epl_d2q9}. By relaxing the moment $k_i$ with a frequency $\omega_i$, the collision operator reads as follows
\begin{equation} \label{collision}
k_i^{\star} = k_i+\omega_i \left(k_i^{eq}-k_i\right), \quad \mathrm{with}\,\, i=3 \ldots 8.
\end{equation}
Only the frequencies related to $k_4$ and $k_5$, \textit{i.e.} $\omega_4$ and $\omega_5$, are linked to the fluid kinematic viscosity, \textit{i.e.} $\omega_{\rho} = \omega_4 = \omega_5$, and the frequency $\omega_3$ is related to the bulk viscosity. Let us collect post-collision central moments and populations as $\displaystyle | k_i^{\star} \rangle = \left[ \rho,\,0,\,0, \, k_3^{\star},\, \ldots, \, k_{8}^{\star}   \right]^{\top}$ and $\displaystyle | f_i^{\star} \rangle = \left[f_0^{\star}, \ldots, f_{8}^{\star} \right]^{\top}$, respectively. The latter are available through $\displaystyle | f_i^{\star} \rangle = \left(\bar{\mathcal{T}}^{\top}\right)^{-1} | k_i^{\star} \rangle$ and, eventually, are streamed. For the sake of computational efficiency, it is worth to stress that only few pre-collision CMs should be computed. In fact, $k_0=\rho,\, k_1=0$ and $k_2=0$ are invariant with respect to the collision. Moreover, $k_6,\, k_7$ and $k_8$ are not involved in the computation because the frequencies $\omega_6$, $\omega_7$ and  $\omega_8$ are associated to higher-order moments and are set equal to 1 in order to enhance the stability of the algorithm, \textit{i.e.} $k_6^{\star}=k_6^{eq},\, k_7^{\star}=k_7^{eq}$ and $k_8^{\star}=k_8^{eq}$. In the Supplementary Material, a script \footnote{See Supplemental Material at [\protect \url{D2Q9_CentralMoments.m}] for performing all the computations to obtain $k_i$, $k_i^{eq}$ and $f_i^{\star}$} allows the reader to perform the symbolic manipulations to obtain all the involved quantities.\\
\indent Now, we rerun the same above-described case with the proposed method. We find that the stability is greatly enhanced. In fact, no apparent limit to the viscosity ratio is achieved, as the model does not show any instability in the limit of $\nu_g \rightarrow 0$. In Fig.~\ref{Figure2}, the velocity map is reported at $\zeta=100$. It is possible to appreciate the smoothness of the solution for a viscosity ratio that is considerably higher than the experienced limit for the double-BGK run.\\
\indent With this set-up, we test the ability of our model to reproduce Laplace's law, stating the surface tension $\sigma$ should be constant or, in other words, the pressure jump across the bubble interface should be proportional to $R^{-1}$. We perform several analyzes by modifying the bubble radius. Moreover, the viscosity ratio varies as $\zeta=[2,10, 100, 1000]$. In Fig.~\ref{Figure3}, the pressure jump $\Delta P$ is plotted against the inverse of the radius for different values of $\zeta$. Consistently with the analytical predictions, findings are not sensitive to the viscosity ratio. Moreover, these appear to be well-distributed along a line whose slope is 0.03717. Interestingly, the theoretical value is 0.03771 (with $\gamma=a=0.04$) and a very satisfactory relative error of $1.43\%$ is achieved.
\\
\indent A second test involves contact angles, which are simulated by the change of the wall gradient $\partial_{\perp}\phi_{w}$ in the direction normal to a surface. These may be predicted by solving the equation
\begin{equation}\label{angles}
\partial_{\perp}\phi_{w} = \sqrt{\frac{2a}{\gamma}   \cos \left(  \frac{\vartheta}{3} \right) \left(  1-   \cos \left(  \frac{\vartheta}{3} \right)\right)},
\end{equation}
where $\vartheta$ is connected to the contact angle as $\vartheta= \cos^{-1} \left(  \sin^2\left(  \theta \right) \right)$. Let us consider a domain composed by 100 points in each direction and a bubble of radius 20 initially placed in the center of the domain. The order parameter is set to $\phi=-1$ everywhere, except for the area covered by the bubble where $\phi=-1$. Moreover, $\gamma=a=0.04$, $\nu_l=0.1$ and $\zeta=2$. Due to the presence of a constant uniform vertical downward force of magnitude $1 \times 10^{-5}$, the bubble impacts the bottom wall where different wall gradients are enforced. No-slip conditions are prescribed at the bottom and top walls, while periodicity is applied to the leftmost and rightmost section of the domain. We perform several analyzes with different values of $\partial_{\perp}\phi_{w}$ by adopting the double-BGK and hybrid CMs-BGK schemes and measure the relative difference $\varepsilon$ between analytical predictions (see Eq.~(\ref{angles})) and values from our numerical runs. Findings are plotted in Fig.~\ref{Figure4}, demonstrating a good agreement. The hybrid scheme exhibits errors that decrease as $\partial_{\perp}\phi_{wall}$ reduces, that is fully consistent with findings from BGK runs.
The properties of the hybrid scheme can be appreciated in Fig.~\ref{Figure5} and \ref{Figure6}, where the different shapes of the interface are plotted for several values of the wall gradient. The bubble shows hydrophobic and hydrophilic behaviors for negative and positive values of $\partial_{\perp}\phi_{w}$, respectively.
\\
\indent Finally, we consider the deformation of a droplet in a shear flow. The domain consists of $2H=200$ points in each direction. A bubble of radius $R=30$ is placed in the center of the fluid domain. A constant uniform horizontal velocity $v_w$ is applied in the rightward direction to the top wall and in the leftward one to the bottom wall. Periodic boundary conditions are enforced at the vertical sides of the domain. The viscosity is $\nu_l = \nu_g=0.2$. The problem is governed by the Reynolds number $\displaystyle \mathrm{Re}=\frac{v_wR^2}{H\nu_l}$ and the capillarity number $\displaystyle \mathrm{Ca}=\frac{v_w R \nu_l \rho}{H\sigma}$. According to Taylor~\cite{taylor1932viscosity}, the deformation $D_f$ of the droplet is proportional to $\displaystyle \frac{35 \mathrm{Ca}}{32}=1.09375 \mathrm{Ca}$ in the Stokes regime (\textit{i.e.},$\mathrm{Re} \ll 1$). The deformation is computed as $\displaystyle D_f = \frac{A-B}{A+B}$, $A$ and $B$ being the lengths of the major and minor axes of the deformed droplet, respectively. By varying the capillarity number, we perform several runs at $\displaystyle \mathrm{Re}=0.1$. The droplet deformation is plotted against Ca in Fig.~\ref{Figure7}. Our findings show a slight mismatch with respect to the analytical references values, as the slope of the fitting line is 1.09536 and the relative discrepancy between the two is highly satisfactory ($0.15\%$).
In Fig.~\ref{Figure8}, the shape of the interface is sketched for different values of capillarity number. As Ca increases (\textit{i.e.}, the viscous force of droplets grows) the degree of deformation slightly increases too; however, it is found that Ca does not strongly affect the droplet's final shape.
\\
\indent We also prove that our model is able to simulate the droplet breakup~\cite{liu2012three}. Let us consider a domain of $240 \times 60$ points with a bubble of radius equal to 15 placed in the center of the domain. This case shares the remaining model parameters with the previous setup. Let us define a characteristic time $t^{\dagger} = H/v_w$, where $H=30$. In Fig.~\ref{Figure10}, the map of the order parameter is plotted at salient time instants. The bubble progressively elongates, while remaining continuous until $t \sim 20 t^{\dagger}$. At $t \sim 25 t^{\dagger}$, the two extreme portions separate from the rest of the body. Where the central part shows a very small thickness, a second breakup is experienced, leading to the rise of four sub-droplets into the systems (\textit{i.e.} the so-called satellite droplets~\cite{Stone1994}). Droplet breakup behavior with such satellite structures can be seen in previous works carried out with the VOF method~\cite{Li2000} and the LB method~\cite{Inamuro2003, liu2012three, Komrakova2014}. The present central-moments-based scheme can also reproduce such complex droplet dynamics.
\\
\indent Here, we have demonstrated that the solution of the coupled Cahn-Hilliard-Navier-Stokes equations can be recovered by a double-populations hybrid CMs-BGK scheme. Numerical tests showed excellent accuracy. Moreover, it allows us to drastically extend the range of simulated viscosity ratio with respect to the double-populations double-BGK method.

\section*{Acknowledgments}
This article is based upon work from COST Action MP1305, supported by COST (European Cooperation in Science and Technology).\\
The support of JSPS KAKENHI Grant Number 16J02077 is also acknowledged.

\bibliography{bibliography}

\begin{thebibliography}{45}%
\makeatletter
\providecommand \@ifxundefined [1]{%
 \@ifx{#1\undefined}
}%
\providecommand \@ifnum [1]{%
 \ifnum #1\expandafter \@firstoftwo
 \else \expandafter \@secondoftwo
 \fi
}%
\providecommand \@ifx [1]{%
 \ifx #1\expandafter \@firstoftwo
 \else \expandafter \@secondoftwo
 \fi
}%
\providecommand \natexlab [1]{#1}%
\providecommand \enquote  [1]{``#1''}%
\providecommand \bibnamefont  [1]{#1}%
\providecommand \bibfnamefont [1]{#1}%
\providecommand \citenamefont [1]{#1}%
\providecommand \href@noop [0]{\@secondoftwo}%
\providecommand \href [0]{\begingroup \@sanitize@url \@href}%
\providecommand \@href[1]{\@@startlink{#1}\@@href}%
\providecommand \@@href[1]{\endgroup#1\@@endlink}%
\providecommand \@sanitize@url [0]{\catcode `\\12\catcode `\$12\catcode
  `\&12\catcode `\#12\catcode `\^12\catcode `\_12\catcode `\%12\relax}%
\providecommand \@@startlink[1]{}%
\providecommand \@@endlink[0]{}%
\providecommand \url  [0]{\begingroup\@sanitize@url \@url }%
\providecommand \@url [1]{\endgroup\@href {#1}{\urlprefix }}%
\providecommand \urlprefix  [0]{URL }%
\providecommand \Eprint [0]{\href }%
\providecommand \doibase [0]{http://dx.doi.org/}%
\providecommand \selectlanguage [0]{\@gobble}%
\providecommand \bibinfo  [0]{\@secondoftwo}%
\providecommand \bibfield  [0]{\@secondoftwo}%
\providecommand \translation [1]{[#1]}%
\providecommand \BibitemOpen [0]{}%
\providecommand \bibitemStop [0]{}%
\providecommand \bibitemNoStop [0]{.\EOS\space}%
\providecommand \EOS [0]{\spacefactor3000\relax}%
\providecommand \BibitemShut  [1]{\csname bibitem#1\endcsname}%
\let\auto@bib@innerbib\@empty
\bibitem [{\citenamefont {van~der Waals}(1979)}]{van1979thermodynamic}%
  \BibitemOpen
  \bibfield  {author} {\bibinfo {author} {\bibfnamefont {J.~D.}\ \bibnamefont
  {van~der Waals}},\ }\href@noop {} {\bibfield  {journal} {\bibinfo  {journal}
  {J. Stat. Phys.}\ }\textbf {\bibinfo {volume} {20}},\ \bibinfo {pages} {200}
  (\bibinfo {year} {1979})}\BibitemShut {NoStop}%
\bibitem [{\citenamefont {Cahn}\ and\ \citenamefont
  {Hilliard}(1958)}]{cahn1958free}%
  \BibitemOpen
  \bibfield  {author} {\bibinfo {author} {\bibfnamefont {J.~W.}\ \bibnamefont
  {Cahn}}\ and\ \bibinfo {author} {\bibfnamefont {J.~E.}\ \bibnamefont
  {Hilliard}},\ }\href@noop {} {\bibfield  {journal} {\bibinfo  {journal} {J.
  Chem. Phys.}\ }\textbf {\bibinfo {volume} {28}},\ \bibinfo {pages} {258}
  (\bibinfo {year} {1958})}\BibitemShut {NoStop}%
\bibitem [{\citenamefont {Cahn}\ and\ \citenamefont
  {Hilliard}(1959)}]{cahn1959free}%
  \BibitemOpen
  \bibfield  {author} {\bibinfo {author} {\bibfnamefont {J.~W.}\ \bibnamefont
  {Cahn}}\ and\ \bibinfo {author} {\bibfnamefont {J.~E.}\ \bibnamefont
  {Hilliard}},\ }\href@noop {} {\bibfield  {journal} {\bibinfo  {journal} {J.
  Chem. Phys.}\ }\textbf {\bibinfo {volume} {31}},\ \bibinfo {pages} {688}
  (\bibinfo {year} {1959})}\BibitemShut {NoStop}%
\bibitem [{\citenamefont {Penrose}\ and\ \citenamefont
  {Fife}(1990)}]{penrose1990thermodynamically}%
  \BibitemOpen
  \bibfield  {author} {\bibinfo {author} {\bibfnamefont {O.}~\bibnamefont
  {Penrose}}\ and\ \bibinfo {author} {\bibfnamefont {P.~C.}\ \bibnamefont
  {Fife}},\ }\href@noop {} {\bibfield  {journal} {\bibinfo  {journal} {Physica
  D}\ }\textbf {\bibinfo {volume} {43}},\ \bibinfo {pages} {44} (\bibinfo
  {year} {1990})}\BibitemShut {NoStop}%
\bibitem [{\citenamefont {Badalassi}\ \emph {et~al.}(2003)\citenamefont
  {Badalassi}, \citenamefont {Ceniceros},\ and\ \citenamefont
  {Banerjee}}]{badalassi2003computation}%
  \BibitemOpen
  \bibfield  {author} {\bibinfo {author} {\bibfnamefont {V.}~\bibnamefont
  {Badalassi}}, \bibinfo {author} {\bibfnamefont {H.}~\bibnamefont
  {Ceniceros}}, \ and\ \bibinfo {author} {\bibfnamefont {S.}~\bibnamefont
  {Banerjee}},\ }\href@noop {} {\bibfield  {journal} {\bibinfo  {journal} {J.
  Comput. Phys.}\ }\textbf {\bibinfo {volume} {190}},\ \bibinfo {pages} {371}
  (\bibinfo {year} {2003})}\BibitemShut {NoStop}%
\bibitem [{\citenamefont {Jacqmin}(1999)}]{jacqmin1999calculation}%
  \BibitemOpen
  \bibfield  {author} {\bibinfo {author} {\bibfnamefont {D.}~\bibnamefont
  {Jacqmin}},\ }\href@noop {} {\bibfield  {journal} {\bibinfo  {journal} {J.
  Comput. Phys.}\ }\textbf {\bibinfo {volume} {155}},\ \bibinfo {pages} {96}
  (\bibinfo {year} {1999})}\BibitemShut {NoStop}%
\bibitem [{\citenamefont {Kim}(2005)}]{kim2005continuous}%
  \BibitemOpen
  \bibfield  {author} {\bibinfo {author} {\bibfnamefont {J.}~\bibnamefont
  {Kim}},\ }\href@noop {} {\bibfield  {journal} {\bibinfo  {journal} {J.
  Comput. Phys.}\ }\textbf {\bibinfo {volume} {204}},\ \bibinfo {pages} {784}
  (\bibinfo {year} {2005})}\BibitemShut {NoStop}%
\bibitem [{\citenamefont {Benzi}\ \emph {et~al.}(1992)\citenamefont {Benzi},
  \citenamefont {Succi},\ and\ \citenamefont {Vergassola}}]{benzi1992lattice}%
  \BibitemOpen
  \bibfield  {author} {\bibinfo {author} {\bibfnamefont {R.}~\bibnamefont
  {Benzi}}, \bibinfo {author} {\bibfnamefont {S.}~\bibnamefont {Succi}}, \ and\
  \bibinfo {author} {\bibfnamefont {M.}~\bibnamefont {Vergassola}},\
  }\href@noop {} {\bibfield  {journal} {\bibinfo  {journal} {Phys. Rep.}\
  }\textbf {\bibinfo {volume} {222}},\ \bibinfo {pages} {145} (\bibinfo {year}
  {1992})}\BibitemShut {NoStop}%
\bibitem [{\citenamefont {He}\ and\ \citenamefont {Luo}(1997)}]{he1997theory}%
  \BibitemOpen
  \bibfield  {author} {\bibinfo {author} {\bibfnamefont {X.}~\bibnamefont
  {He}}\ and\ \bibinfo {author} {\bibfnamefont {L.-S.}\ \bibnamefont {Luo}},\
  }\href@noop {} {\bibfield  {journal} {\bibinfo  {journal} {Phys. Rev. E}\
  }\textbf {\bibinfo {volume} {56}},\ \bibinfo {pages} {6811} (\bibinfo {year}
  {1997})}\BibitemShut {NoStop}%
\bibitem [{\citenamefont {Chen}\ and\ \citenamefont
  {Doolen}(1998)}]{chen1998lattice}%
  \BibitemOpen
  \bibfield  {author} {\bibinfo {author} {\bibfnamefont {S.}~\bibnamefont
  {Chen}}\ and\ \bibinfo {author} {\bibfnamefont {G.~D.}\ \bibnamefont
  {Doolen}},\ }\href@noop {} {\bibfield  {journal} {\bibinfo  {journal} {Annu.
  Rev. Fluid Mech.}\ }\textbf {\bibinfo {volume} {30}},\ \bibinfo {pages} {329}
  (\bibinfo {year} {1998})}\BibitemShut {NoStop}%
\bibitem [{\citenamefont {He}\ \emph {et~al.}(1999)\citenamefont {He},
  \citenamefont {Chen},\ and\ \citenamefont {Zhang}}]{he1999lattice}%
  \BibitemOpen
  \bibfield  {author} {\bibinfo {author} {\bibfnamefont {X.}~\bibnamefont
  {He}}, \bibinfo {author} {\bibfnamefont {S.}~\bibnamefont {Chen}}, \ and\
  \bibinfo {author} {\bibfnamefont {R.}~\bibnamefont {Zhang}},\ }\href@noop {}
  {\bibfield  {journal} {\bibinfo  {journal} {J. Comput. Phys.}\ }\textbf
  {\bibinfo {volume} {152}},\ \bibinfo {pages} {642} (\bibinfo {year}
  {1999})}\BibitemShut {NoStop}%
\bibitem [{\citenamefont {Briant}\ and\ \citenamefont
  {Yeomans}(2004)}]{briant2004lattice}%
  \BibitemOpen
  \bibfield  {author} {\bibinfo {author} {\bibfnamefont {A.}~\bibnamefont
  {Briant}}\ and\ \bibinfo {author} {\bibfnamefont {J.}~\bibnamefont
  {Yeomans}},\ }\href@noop {} {\bibfield  {journal} {\bibinfo  {journal} {Phys.
  Rev. E}\ }\textbf {\bibinfo {volume} {69}},\ \bibinfo {pages} {031603}
  (\bibinfo {year} {2004})}\BibitemShut {NoStop}%
\bibitem [{\citenamefont {Inamuro}\ \emph {et~al.}(2004)\citenamefont
  {Inamuro}, \citenamefont {Ogata}, \citenamefont {Tajima},\ and\ \citenamefont
  {Konishi}}]{Inamuro2004}%
  \BibitemOpen
  \bibfield  {author} {\bibinfo {author} {\bibfnamefont {T.}~\bibnamefont
  {Inamuro}}, \bibinfo {author} {\bibfnamefont {T.}~\bibnamefont {Ogata}},
  \bibinfo {author} {\bibfnamefont {S.}~\bibnamefont {Tajima}}, \ and\ \bibinfo
  {author} {\bibfnamefont {N.}~\bibnamefont {Konishi}},\ }\href@noop {}
  {\bibfield  {journal} {\bibinfo  {journal} {J. Comput. Phys.}\ }\textbf
  {\bibinfo {volume} {198}},\ \bibinfo {pages} {628} (\bibinfo {year}
  {2004})}\BibitemShut {NoStop}%
\bibitem [{\citenamefont {Lee}\ and\ \citenamefont {Lin}(2005)}]{Lee2005}%
  \BibitemOpen
  \bibfield  {author} {\bibinfo {author} {\bibfnamefont {T.}~\bibnamefont
  {Lee}}\ and\ \bibinfo {author} {\bibfnamefont {C.-L.}\ \bibnamefont {Lin}},\
  }\href {\doibase 10.1016/j.jcp.2004.12.001} {\bibfield  {journal} {\bibinfo
  {journal} {J. Comput. Phys.}\ }\textbf {\bibinfo {volume} {206}},\ \bibinfo
  {pages} {16} (\bibinfo {year} {2005})}\BibitemShut {NoStop}%
\bibitem [{\citenamefont {Zheng}\ \emph {et~al.}(2006)\citenamefont {Zheng},
  \citenamefont {Shu},\ and\ \citenamefont {Chew}}]{Zheng2006}%
  \BibitemOpen
  \bibfield  {author} {\bibinfo {author} {\bibfnamefont {H.~W.}\ \bibnamefont
  {Zheng}}, \bibinfo {author} {\bibfnamefont {C.}~\bibnamefont {Shu}}, \ and\
  \bibinfo {author} {\bibfnamefont {Y.~T.}\ \bibnamefont {Chew}},\ }\href
  {\doibase 10.1016/j.jcp.2006.02.015} {\bibfield  {journal} {\bibinfo
  {journal} {J. Comput. Phys.}\ }\textbf {\bibinfo {volume} {218}},\ \bibinfo
  {pages} {353} (\bibinfo {year} {2006})}\BibitemShut {NoStop}%
\bibitem [{\citenamefont {{Mazloomi}}\ \emph {et~al.}(2015)\citenamefont
  {{Mazloomi}}, \citenamefont {Chikatamarla},\ and\ \citenamefont
  {Karlin}}]{MazloomiM2015}%
  \BibitemOpen
  \bibfield  {author} {\bibinfo {author} {\bibfnamefont {M.~A.}\ \bibnamefont
  {{Mazloomi}}}, \bibinfo {author} {\bibfnamefont {S.~S.}\ \bibnamefont
  {Chikatamarla}}, \ and\ \bibinfo {author} {\bibfnamefont {I.~V.}\
  \bibnamefont {Karlin}},\ }\href {\doibase 10.1103/PhysRevLett.114.174502}
  {\bibfield  {journal} {\bibinfo  {journal} {Phys. Rev. Lett.}\ }\textbf
  {\bibinfo {volume} {114}},\ \bibinfo {pages} {174502} (\bibinfo {year}
  {2015})}\BibitemShut {NoStop}%
\bibitem [{\citenamefont {Zheng}\ \emph {et~al.}(2005)\citenamefont {Zheng},
  \citenamefont {Shu},\ and\ \citenamefont {Chew}}]{zheng2005lattice}%
  \BibitemOpen
  \bibfield  {author} {\bibinfo {author} {\bibfnamefont {H.}~\bibnamefont
  {Zheng}}, \bibinfo {author} {\bibfnamefont {C.}~\bibnamefont {Shu}}, \ and\
  \bibinfo {author} {\bibfnamefont {Y.}~\bibnamefont {Chew}},\ }\href@noop {}
  {\bibfield  {journal} {\bibinfo  {journal} {Phys. Rev. E}\ }\textbf {\bibinfo
  {volume} {72}},\ \bibinfo {pages} {056705} (\bibinfo {year}
  {2005})}\BibitemShut {NoStop}%
\bibitem [{\citenamefont {Lee}\ and\ \citenamefont
  {Liu}(2010)}]{lee2010lattice}%
  \BibitemOpen
  \bibfield  {author} {\bibinfo {author} {\bibfnamefont {T.}~\bibnamefont
  {Lee}}\ and\ \bibinfo {author} {\bibfnamefont {L.}~\bibnamefont {Liu}},\
  }\href@noop {} {\bibfield  {journal} {\bibinfo  {journal} {J. Comput. Phys.}\
  }\textbf {\bibinfo {volume} {229}},\ \bibinfo {pages} {8045} (\bibinfo {year}
  {2010})}\BibitemShut {NoStop}%
\bibitem [{\citenamefont {Zu}\ and\ \citenamefont {He}(2013)}]{zu2013phase}%
  \BibitemOpen
  \bibfield  {author} {\bibinfo {author} {\bibfnamefont {Y.}~\bibnamefont
  {Zu}}\ and\ \bibinfo {author} {\bibfnamefont {S.}~\bibnamefont {He}},\
  }\href@noop {} {\bibfield  {journal} {\bibinfo  {journal} {Phys.l Rev. E}\
  }\textbf {\bibinfo {volume} {87}},\ \bibinfo {pages} {043301} (\bibinfo
  {year} {2013})}\BibitemShut {NoStop}%
\bibitem [{\citenamefont {Zheng}\ \emph {et~al.}(2015)\citenamefont {Zheng},
  \citenamefont {Zheng},\ and\ \citenamefont {Zhai}}]{zheng2015lattice}%
  \BibitemOpen
  \bibfield  {author} {\bibinfo {author} {\bibfnamefont {L.}~\bibnamefont
  {Zheng}}, \bibinfo {author} {\bibfnamefont {S.}~\bibnamefont {Zheng}}, \ and\
  \bibinfo {author} {\bibfnamefont {Q.}~\bibnamefont {Zhai}},\ }\href@noop {}
  {\bibfield  {journal} {\bibinfo  {journal} {Physical Review E}\ }\textbf
  {\bibinfo {volume} {91}},\ \bibinfo {pages} {013309} (\bibinfo {year}
  {2015})}\BibitemShut {NoStop}%
\bibitem [{\citenamefont {Li}\ \emph {et~al.}(2016)\citenamefont {Li},
  \citenamefont {Luo}, \citenamefont {Kang}, \citenamefont {He}, \citenamefont
  {Chen},\ and\ \citenamefont {Liu}}]{li2016lattice}%
  \BibitemOpen
  \bibfield  {author} {\bibinfo {author} {\bibfnamefont {Q.}~\bibnamefont
  {Li}}, \bibinfo {author} {\bibfnamefont {K.}~\bibnamefont {Luo}}, \bibinfo
  {author} {\bibfnamefont {Q.}~\bibnamefont {Kang}}, \bibinfo {author}
  {\bibfnamefont {Y.}~\bibnamefont {He}}, \bibinfo {author} {\bibfnamefont
  {Q.}~\bibnamefont {Chen}}, \ and\ \bibinfo {author} {\bibfnamefont
  {Q.}~\bibnamefont {Liu}},\ }\href@noop {} {\bibfield  {journal} {\bibinfo
  {journal} {Prog. Energ. Combust.}\ }\textbf {\bibinfo {volume} {52}},\
  \bibinfo {pages} {62} (\bibinfo {year} {2016})}\BibitemShut {NoStop}%
\bibitem [{\citenamefont {Succi}(2001)}]{SucciBook}%
  \BibitemOpen
  \bibfield  {author} {\bibinfo {author} {\bibfnamefont {S.}~\bibnamefont
  {Succi}},\ }\href@noop {} {\emph {\bibinfo {title} {The Lattice {B}oltzmann
  Equation for Fluid Dynamics and Beyond}}}\ (\bibinfo  {publisher}
  {Clarendon},\ \bibinfo {year} {2001})\BibitemShut {NoStop}%
\bibitem [{\citenamefont {Bhatnagar}\ \emph {et~al.}(1954)\citenamefont
  {Bhatnagar}, \citenamefont {Gross},\ and\ \citenamefont
  {Krook}}]{bhatnagar1954model}%
  \BibitemOpen
  \bibfield  {author} {\bibinfo {author} {\bibfnamefont {P.}~\bibnamefont
  {Bhatnagar}}, \bibinfo {author} {\bibfnamefont {E.}~\bibnamefont {Gross}}, \
  and\ \bibinfo {author} {\bibfnamefont {M.}~\bibnamefont {Krook}},\
  }\href@noop {} {\bibfield  {journal} {\bibinfo  {journal} {Phys. Rev.}\
  }\textbf {\bibinfo {volume} {94}},\ \bibinfo {pages} {511} (\bibinfo {year}
  {1954})}\BibitemShut {NoStop}%
\bibitem [{\citenamefont {Pooley}\ and\ \citenamefont
  {Furtado}(2008)}]{pooley2008eliminating}%
  \BibitemOpen
  \bibfield  {author} {\bibinfo {author} {\bibfnamefont {C.}~\bibnamefont
  {Pooley}}\ and\ \bibinfo {author} {\bibfnamefont {K.}~\bibnamefont
  {Furtado}},\ }\href@noop {} {\bibfield  {journal} {\bibinfo  {journal} {Phys.
  Rev. E}\ }\textbf {\bibinfo {volume} {77}},\ \bibinfo {pages} {046702}
  (\bibinfo {year} {2008})}\BibitemShut {NoStop}%
\bibitem [{\citenamefont {Inamuro}\ \emph {et~al.}(2000)\citenamefont
  {Inamuro}, \citenamefont {Konishi},\ and\ \citenamefont
  {Ogino}}]{inamuro2000galilean}%
  \BibitemOpen
  \bibfield  {author} {\bibinfo {author} {\bibfnamefont {T.}~\bibnamefont
  {Inamuro}}, \bibinfo {author} {\bibfnamefont {N.}~\bibnamefont {Konishi}}, \
  and\ \bibinfo {author} {\bibfnamefont {F.}~\bibnamefont {Ogino}},\
  }\href@noop {} {\bibfield  {journal} {\bibinfo  {journal} {Comput. Phys.
  Comm.}\ }\textbf {\bibinfo {volume} {129}},\ \bibinfo {pages} {32} (\bibinfo
  {year} {2000})}\BibitemShut {NoStop}%
\bibitem [{\citenamefont {Kalarakis}\ \emph {et~al.}(2002)\citenamefont
  {Kalarakis}, \citenamefont {Burganos},\ and\ \citenamefont
  {Payatakes}}]{kalarakis2002galilean}%
  \BibitemOpen
  \bibfield  {author} {\bibinfo {author} {\bibfnamefont {A.}~\bibnamefont
  {Kalarakis}}, \bibinfo {author} {\bibfnamefont {V.}~\bibnamefont {Burganos}},
  \ and\ \bibinfo {author} {\bibfnamefont {A.}~\bibnamefont {Payatakes}},\
  }\href@noop {} {\bibfield  {journal} {\bibinfo  {journal} {Phys. Rev. E}\
  }\textbf {\bibinfo {volume} {65}},\ \bibinfo {pages} {056702} (\bibinfo
  {year} {2002})}\BibitemShut {NoStop}%
\bibitem [{\citenamefont {Pooley}\ \emph {et~al.}(2008)\citenamefont {Pooley},
  \citenamefont {Kusumaatmaja},\ and\ \citenamefont
  {Yeomans}}]{pooley2008contact}%
  \BibitemOpen
  \bibfield  {author} {\bibinfo {author} {\bibfnamefont {C.}~\bibnamefont
  {Pooley}}, \bibinfo {author} {\bibfnamefont {H.}~\bibnamefont
  {Kusumaatmaja}}, \ and\ \bibinfo {author} {\bibfnamefont {J.}~\bibnamefont
  {Yeomans}},\ }\href@noop {} {\bibfield  {journal} {\bibinfo  {journal} {Phys.
  Rev. E}\ }\textbf {\bibinfo {volume} {78}},\ \bibinfo {pages} {056709}
  (\bibinfo {year} {2008})}\BibitemShut {NoStop}%
\bibitem [{\citenamefont {Liang}\ \emph {et~al.}(2014)\citenamefont {Liang},
  \citenamefont {Shi}, \citenamefont {Guo},\ and\ \citenamefont
  {Chai}}]{liang2014phase}%
  \BibitemOpen
  \bibfield  {author} {\bibinfo {author} {\bibfnamefont {H.}~\bibnamefont
  {Liang}}, \bibinfo {author} {\bibfnamefont {B.}~\bibnamefont {Shi}}, \bibinfo
  {author} {\bibfnamefont {Z.}~\bibnamefont {Guo}}, \ and\ \bibinfo {author}
  {\bibfnamefont {Z.}~\bibnamefont {Chai}},\ }\href@noop {} {\bibfield
  {journal} {\bibinfo  {journal} {Phys. Rev. E}\ }\textbf {\bibinfo {volume}
  {89}},\ \bibinfo {pages} {053320} (\bibinfo {year} {2014})}\BibitemShut
  {NoStop}%
\bibitem [{\citenamefont {Geier}\ \emph {et~al.}(2006)\citenamefont {Geier},
  \citenamefont {Greiner},\ and\ \citenamefont {Korvink}}]{geier2006cascaded}%
  \BibitemOpen
  \bibfield  {author} {\bibinfo {author} {\bibfnamefont {M.}~\bibnamefont
  {Geier}}, \bibinfo {author} {\bibfnamefont {A.}~\bibnamefont {Greiner}}, \
  and\ \bibinfo {author} {\bibfnamefont {J.}~\bibnamefont {Korvink}},\
  }\href@noop {} {\bibfield  {journal} {\bibinfo  {journal} {Phys. Rev. E}\
  }\textbf {\bibinfo {volume} {73}},\ \bibinfo {pages} {066705} (\bibinfo
  {year} {2006})}\BibitemShut {NoStop}%
\bibitem [{\citenamefont {Premnath}\ and\ \citenamefont
  {Banerjee}(2009)}]{premnath2009incorporating}%
  \BibitemOpen
  \bibfield  {author} {\bibinfo {author} {\bibfnamefont {K.}~\bibnamefont
  {Premnath}}\ and\ \bibinfo {author} {\bibfnamefont {S.}~\bibnamefont
  {Banerjee}},\ }\href@noop {} {\bibfield  {journal} {\bibinfo  {journal}
  {Phys. Rev. E}\ }\textbf {\bibinfo {volume} {80}},\ \bibinfo {pages} {036702}
  (\bibinfo {year} {2009})}\BibitemShut {NoStop}%
\bibitem [{\citenamefont {Ning}\ \emph {et~al.}(2016)\citenamefont {Ning},
  \citenamefont {Premnath},\ and\ \citenamefont {Patil}}]{FLD:FLD4208}%
  \BibitemOpen
  \bibfield  {author} {\bibinfo {author} {\bibfnamefont {Y.}~\bibnamefont
  {Ning}}, \bibinfo {author} {\bibfnamefont {K.~N.}\ \bibnamefont {Premnath}},
  \ and\ \bibinfo {author} {\bibfnamefont {D.~V.}\ \bibnamefont {Patil}},\
  }\href@noop {} {\bibfield  {journal} {\bibinfo  {journal} {Int. J. Numer.
  Meth. Fl.}\ }\textbf {\bibinfo {volume} {82}},\ \bibinfo {pages} {59}
  (\bibinfo {year} {2016})}\BibitemShut {NoStop}%
\bibitem [{\citenamefont {Geier}\ \emph
  {et~al.}(2015{\natexlab{a}})\citenamefont {Geier}, \citenamefont
  {Sch{\"o}nherr}, \citenamefont {Pasquali},\ and\ \citenamefont
  {Krafczyk}}]{geier2015cumulant}%
  \BibitemOpen
  \bibfield  {author} {\bibinfo {author} {\bibfnamefont {M.}~\bibnamefont
  {Geier}}, \bibinfo {author} {\bibfnamefont {M.}~\bibnamefont
  {Sch{\"o}nherr}}, \bibinfo {author} {\bibfnamefont {A.}~\bibnamefont
  {Pasquali}}, \ and\ \bibinfo {author} {\bibfnamefont {M.}~\bibnamefont
  {Krafczyk}},\ }\href@noop {} {\bibfield  {journal} {\bibinfo  {journal}
  {Comput. Math. Appl.}\ }\textbf {\bibinfo {volume} {70}},\ \bibinfo {pages}
  {507} (\bibinfo {year} {2015}{\natexlab{a}})}\BibitemShut {NoStop}%
\bibitem [{\citenamefont {Geier}\ \emph
  {et~al.}(2015{\natexlab{b}})\citenamefont {Geier}, \citenamefont {Fakhari},\
  and\ \citenamefont {Lee}}]{geier2015conservative}%
  \BibitemOpen
  \bibfield  {author} {\bibinfo {author} {\bibfnamefont {M.}~\bibnamefont
  {Geier}}, \bibinfo {author} {\bibfnamefont {A.}~\bibnamefont {Fakhari}}, \
  and\ \bibinfo {author} {\bibfnamefont {T.}~\bibnamefont {Lee}},\ }\href@noop
  {} {\bibfield  {journal} {\bibinfo  {journal} {Phys. Rev. E}\ }\textbf
  {\bibinfo {volume} {91}},\ \bibinfo {pages} {063309} (\bibinfo {year}
  {2015}{\natexlab{b}})}\BibitemShut {NoStop}%
\bibitem [{\citenamefont {Fakhari}\ \emph {et~al.}(2016)\citenamefont
  {Fakhari}, \citenamefont {Geier},\ and\ \citenamefont
  {Bolster}}]{fakhari2016simple}%
  \BibitemOpen
  \bibfield  {author} {\bibinfo {author} {\bibfnamefont {A.}~\bibnamefont
  {Fakhari}}, \bibinfo {author} {\bibfnamefont {M.}~\bibnamefont {Geier}}, \
  and\ \bibinfo {author} {\bibfnamefont {D.}~\bibnamefont {Bolster}},\
  }\href@noop {} {\bibfield  {journal} {\bibinfo  {journal} {Comput. Math.
  Appl.}\ } (\bibinfo {year} {2016})}\BibitemShut {NoStop}%
\bibitem [{\citenamefont {De~Rosis}(2016)}]{derosis2016epl_d2q9}%
  \BibitemOpen
  \bibfield  {author} {\bibinfo {author} {\bibfnamefont {A.}~\bibnamefont
  {De~Rosis}},\ }\href@noop {} {\bibfield  {journal} {\bibinfo  {journal}
  {Europhys. Lett.}\ }\textbf {\bibinfo {volume} {116}},\ \bibinfo {pages}
  {44003} (\bibinfo {year} {2016})}\BibitemShut {NoStop}%
\bibitem [{\citenamefont {De~Rosis}(2017{\natexlab{a}})}]{de2017nonorthogonal}%
  \BibitemOpen
  \bibfield  {author} {\bibinfo {author} {\bibfnamefont {A.}~\bibnamefont
  {De~Rosis}},\ }\href@noop {} {\bibfield  {journal} {\bibinfo  {journal}
  {Phys. Rev. E}\ }\textbf {\bibinfo {volume} {95}},\ \bibinfo {pages} {013310}
  (\bibinfo {year} {2017}{\natexlab{a}})}\BibitemShut {NoStop}%
\bibitem [{\citenamefont
  {De~Rosis}(2017{\natexlab{b}})}]{de2017preconditioned}%
  \BibitemOpen
  \bibfield  {author} {\bibinfo {author} {\bibfnamefont {A.}~\bibnamefont
  {De~Rosis}},\ }\href@noop {} {\bibfield  {journal} {\bibinfo  {journal}
  {Phys. Rev. E}\ }\textbf {\bibinfo {volume} {96}},\ \bibinfo {pages} {063308}
  (\bibinfo {year} {2017}{\natexlab{b}})}\BibitemShut {NoStop}%
\bibitem [{\citenamefont
  {De~Rosis}(2017{\natexlab{c}})}]{de2017centralshallow}%
  \BibitemOpen
  \bibfield  {author} {\bibinfo {author} {\bibfnamefont {A.}~\bibnamefont
  {De~Rosis}},\ }\href@noop {} {\bibfield  {journal} {\bibinfo  {journal}
  {Comput. Method Appl. M.}\ }\textbf {\bibinfo {volume} {319}},\ \bibinfo
  {pages} {379} (\bibinfo {year} {2017}{\natexlab{c}})}\BibitemShut {NoStop}%
\bibitem [{Note1()}]{Note1}%
  \BibitemOpen
  \bibinfo {note} {See Supplemental Material at [\protect \url
  {D2Q9_CentralMoments.m}] for performing all the computations to obtain $k_i$,
  $k_i^{eq}$ and $f_i^{\star }$}\BibitemShut {NoStop}%
\bibitem [{\citenamefont {Taylor}(1932)}]{taylor1932viscosity}%
  \BibitemOpen
  \bibfield  {author} {\bibinfo {author} {\bibfnamefont {G.~I.}\ \bibnamefont
  {Taylor}},\ }\href@noop {} {\bibfield  {journal} {\bibinfo  {journal} {Phil.
  Trans. R. Soc. A}\ }\textbf {\bibinfo {volume} {138}},\ \bibinfo {pages} {41}
  (\bibinfo {year} {1932})}\BibitemShut {NoStop}%
\bibitem [{\citenamefont {Liu}\ \emph {et~al.}(2012)\citenamefont {Liu},
  \citenamefont {Valocchi},\ and\ \citenamefont {Kang}}]{liu2012three}%
  \BibitemOpen
  \bibfield  {author} {\bibinfo {author} {\bibfnamefont {H.}~\bibnamefont
  {Liu}}, \bibinfo {author} {\bibfnamefont {A.~J.}\ \bibnamefont {Valocchi}}, \
  and\ \bibinfo {author} {\bibfnamefont {Q.}~\bibnamefont {Kang}},\ }\href@noop
  {} {\bibfield  {journal} {\bibinfo  {journal} {Phys. Rev. E}\ }\textbf
  {\bibinfo {volume} {85}},\ \bibinfo {pages} {046309} (\bibinfo {year}
  {2012})}\BibitemShut {NoStop}%
\bibitem [{\citenamefont {Stone}(1994)}]{Stone1994}%
  \BibitemOpen
  \bibfield  {author} {\bibinfo {author} {\bibfnamefont {H.~A.}\ \bibnamefont
  {Stone}},\ }\href {\doibase 10.1146/annurev.fl.26.010194.000433} {\bibfield
  {journal} {\bibinfo  {journal} {Annu. Rev. Fluid Mech.}\ }\textbf {\bibinfo
  {volume} {26}},\ \bibinfo {pages} {65} (\bibinfo {year} {1994})}\BibitemShut
  {NoStop}%
\bibitem [{\citenamefont {Li}\ \emph {et~al.}(2000)\citenamefont {Li},
  \citenamefont {Renardy},\ and\ \citenamefont {Renardy}}]{Li2000}%
  \BibitemOpen
  \bibfield  {author} {\bibinfo {author} {\bibfnamefont {J.}~\bibnamefont
  {Li}}, \bibinfo {author} {\bibfnamefont {Y.~Y.}\ \bibnamefont {Renardy}}, \
  and\ \bibinfo {author} {\bibfnamefont {M.}~\bibnamefont {Renardy}},\ }\href
  {\doibase 10.1063/1.870305} {\bibfield  {journal} {\bibinfo  {journal} {Phys.
  Fluids}\ }\textbf {\bibinfo {volume} {12}},\ \bibinfo {pages} {269} (\bibinfo
  {year} {2000})}\BibitemShut {NoStop}%
\bibitem [{\citenamefont {Inamuro}\ \emph {et~al.}(2003)\citenamefont
  {Inamuro}, \citenamefont {Tomita},\ and\ \citenamefont
  {Ogino}}]{Inamuro2003}%
  \BibitemOpen
  \bibfield  {author} {\bibinfo {author} {\bibfnamefont {T.}~\bibnamefont
  {Inamuro}}, \bibinfo {author} {\bibfnamefont {R.}~\bibnamefont {Tomita}}, \
  and\ \bibinfo {author} {\bibfnamefont {F.}~\bibnamefont {Ogino}},\ }\href
  {\doibase 10.1142/S0217979203017035} {\bibfield  {journal} {\bibinfo
  {journal} {Int. J. Mod. Phys. B}\ }\textbf {\bibinfo {volume} {17}},\
  \bibinfo {pages} {21} (\bibinfo {year} {2003})}\BibitemShut {NoStop}%
\bibitem [{\citenamefont {Komrakova}\ \emph {et~al.}(2014)\citenamefont
  {Komrakova}, \citenamefont {Shardt}, \citenamefont {Eskin},\ and\
  \citenamefont {Derksen}}]{Komrakova2014}%
  \BibitemOpen
  \bibfield  {author} {\bibinfo {author} {\bibfnamefont {A.}~\bibnamefont
  {Komrakova}}, \bibinfo {author} {\bibfnamefont {O.}~\bibnamefont {Shardt}},
  \bibinfo {author} {\bibfnamefont {D.}~\bibnamefont {Eskin}}, \ and\ \bibinfo
  {author} {\bibfnamefont {J.}~\bibnamefont {Derksen}},\ }\href {\doibase
  10.1016/j.ijmultiphaseflow.2013.10.009} {\bibfield  {journal} {\bibinfo
  {journal} {Int. J. Multiphase Flow}\ }\textbf {\bibinfo {volume} {59}},\
  \bibinfo {pages} {24} (\bibinfo {year} {2014})}\BibitemShut {NoStop}%
\end{thebibliography}%

\end{document}